\documentclass{article}
\usepackage{amssymb,amsmath,amstext,mathtools,ifthen}
\usepackage[pdftex]{graphicx}
\usepackage[pdftex,pagebackref=false]{hyperref}
\usepackage{epsfig}
\usepackage{mathrsfs}
\usepackage{cite}
\usepackage{hyperref}
\usepackage{multirow}
\usepackage{cleveref,setspace,subfig,amsfonts,color,courier}
\usepackage[table]{xcolor}
\numberwithin{equation}{section}

\newcommand{\lambdat}{\widetilde{\lambda}}

\begin{document}

\pagenumbering{roman}
\begin{titlepage}

\begin{center}
{\Large \bf New Variables For Graviton Scattering Amplitudes}
\vspace{0.5cm}

{\bf Yuxiang Gu$^{a,b}$}
\vspace{.1cm}

{\it $^a$ Perimeter Institute for Theoretical Physics, Waterloo, Ontario N2J W29, CA}

{\it $^b$ Physics Department, University of Waterloo, Waterloo, Ontario N2L 3G1, CA}
\end{center}

\vspace{0.5cm}

\begin{abstract}
Motivated by the success of Hodges' momentum twistor variables in planar Yang-Mills,
in this note we introduce a set of new variables, the $\mathcal{S}$ variables, which
are tailored for gravity (or more generally for theories without color ordering). The
$\mathcal{S}$ variables trivialize all on-shell constraints on kinematic data and
momentum conservation while keeping permutation invariance.
We explicitly show the relation between the $\mathcal{S}$ variables and the
spinor-helicity variables $\lambda$ and $\lambdat$ as well as the connection to
momentum twistors. The $\mathcal{S}$ variables can be nicely understood using the
geometry of Grassmannians and are determined by a 2-plane and a 4-plane in
$\mathbb{C}^n$, with $n$ the number of the particles. As an illustration of their
utility, we use the $\mathcal{S}$ variables to present a
reference-free form of soft factors and tree level MHV amplitudes of gravity which is
obtained by using the recent formula \cite{Hodges:2012ym} given by Hodges.
\end{abstract}

\end{titlepage}

\pagenumbering{arabic}
\section{Introduction}
\label{ch:intro}
Recent years have seen huge progress in the computation of scattering amplitudes in
Yang-Mills theory, particularly in the planar limit. In comparison, computing gravity
amplitudes is more difficult.

Among the new challenges, one problem is how to
trivialize momentum conservation in computing gravity amplitudes without obscuring
any symmetries.
In gravity there is no color structure and therefore amplitudes cannot be split into physical
subamplitudes like in Yang-Mills. Maximally supersymmetric gravity amplitudes are
permutation invariant under the exchange of particle labels.
However, imposing momentum conservation in the spinor-helicity variables $\lambda$ and
$\lambdat$ \cite{Xu:1986xb} requires breaking the symmetry.
\begin{equation}
\begin{aligned}
\includegraphics[scale=0.42]{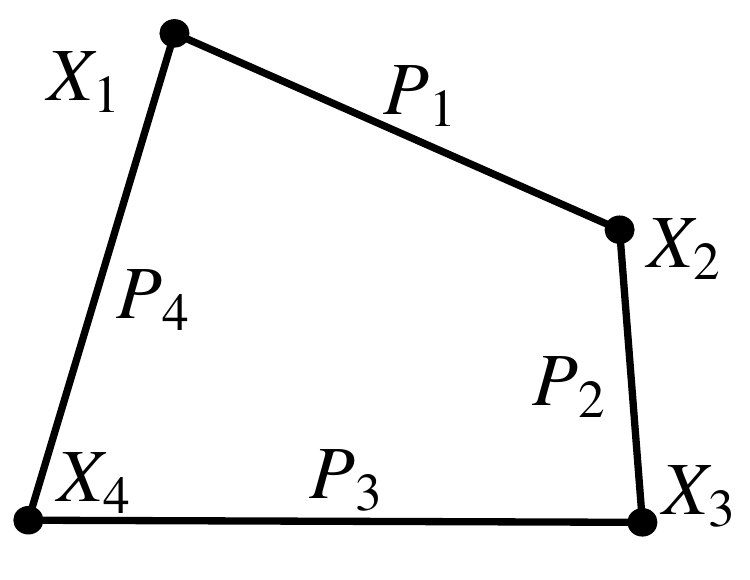} \qquad \qquad  &
\includegraphics[scale=0.32]{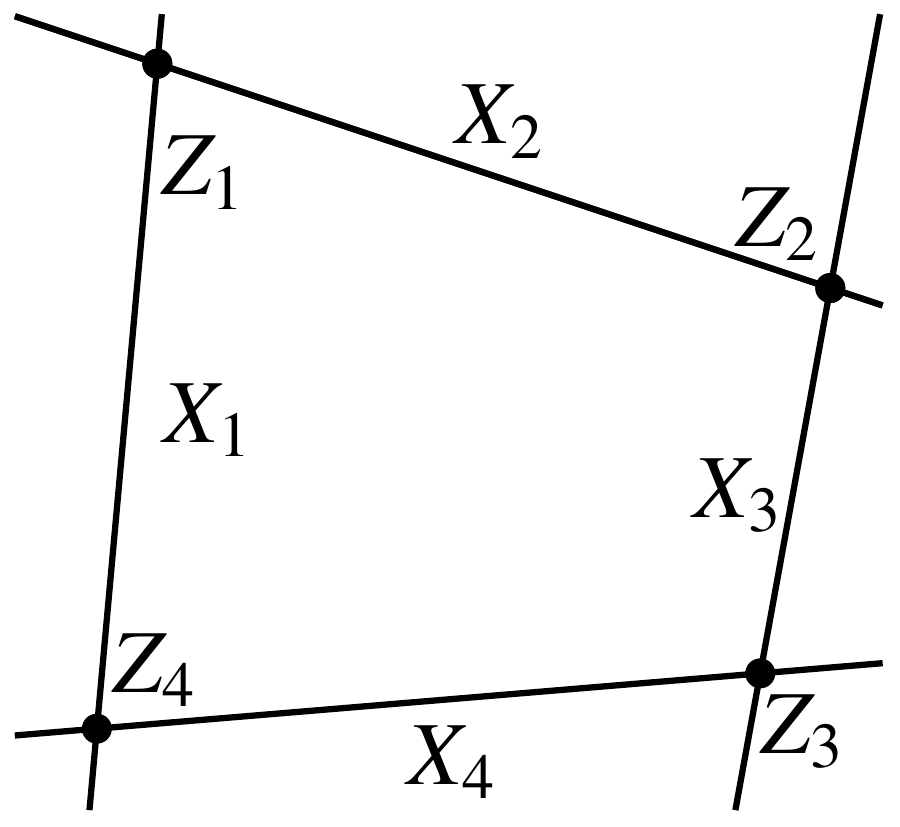}  \\
\text{\small{Dual momentum space}} \qquad \quad \quad   &
  \text{\small{Momentum~twistor~space}}
\end{aligned}
\end{equation}

In planar gauge theory, the same problem was solved in
\cite{Hodges:2009hk} by the introduction of variables called momentum twistors.
Momenta for external particles are defined by a polygonal structure in a dual space as
\cite{Alday:2007hr}
\begin{align}
  \label{eq:momentum_twistor}
{Z_i} =
	\left(\begin{array}{c} \lambda_i \\ \mu_i \end{array} \right),\qquad P_i =
	X_{i+1}-X_i, \qquad X_{ia\dot{a}}\lambda_i^a = \mu _{i\dot{a}}.
\end{align}
In this way the sum of right hand side is always zero so momentum
conservation $\sum_{i=1}^n{P_i}=0$ on the left hand side is satisfied automatically.

\begin{figure}[ht!]
     \centering
     \includegraphics[scale=0.52]{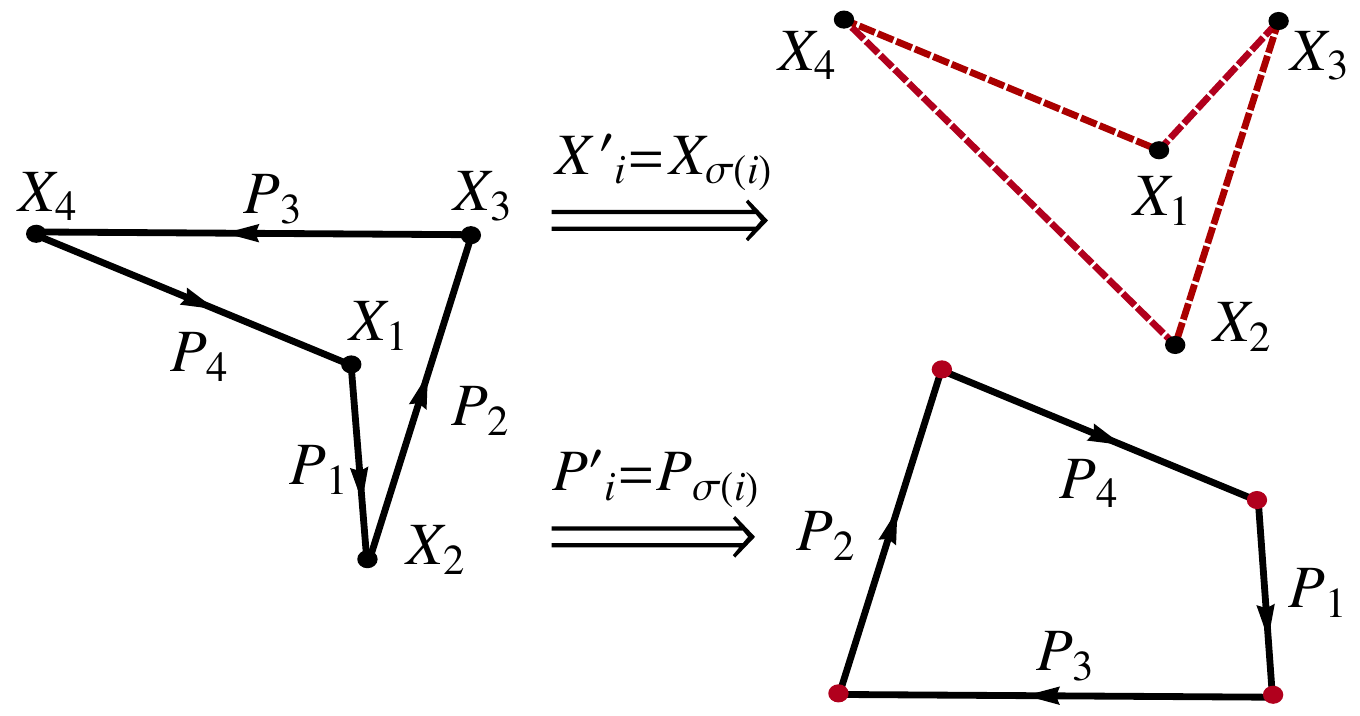}
%      \caption*{}\label{fig:permu_mt}
\end{figure}
However, this construction is not natural in gravity (or nonplanar amplitudes in gauge
theory). The problem lies right in its beauty in the planar sector; it has a
natural ordering in the definition.
This is perfect for objects containing only one fixed ordering, but it might not be very useful
in general.
The figure above shows how a permutation $\sigma$ acting on points $X_i$ in the dual
momentum space is inequivalent to the same permutation acting on momenta $P_i$.
$X_{\sigma(i)}$ violates the original choice of momenta $P_i$, shown as red dashed lines.
$P_{\sigma(i)}$ violates the original $X_i$, shown as red dots. As a result,
a solution to trivializing momentum conservation for gravity amplitudes seems to require a new construction.

In this note we introduce a set of new variables that trivialize momentum conservation
universally including gravity. We call them the $\mathcal{S}$ variables, in which
$\mathcal{S}$ stands for ``symmetric''.  The $\mathcal{S}$ variables consist of a
spinor $\lambda_i$ and a ``twistor"\footnote{We call $Z$ a twistor in a slight abuse
of terminology because it is a 4-component object whose rescaling can be embedded in a
little group transformation and it is closely related to momentum twistors. We show
the relation in
\cref{ch:invitation}.} ${Z}_i$ for each particle $i$. From \cite{ArkaniHamed:2009si}
we know $\lambda$ and $\lambdat$ can be considered as two $2$-dimensional planes in
$\mathbb{C}^n$ which are ``orthogonal"\footnote{Abusing terminology once again, here we
mean that one vector space is in the complement of the other.} to each other. Here in
the $\mathcal{S}$ variables we have the same $2$-plane $\lambda$ and the $SL(2)$
invariants $\langle i~j\rangle$ are the same. Instead of $\lambdat$ we have a
$4$-plane $Z$ and the remaining kinematic building block $[i~j]$ can be shown to be
\begin{equation*}
\begin{aligned}
[i~j]= \sum_{k,l=1}^{n}{\langle
k~l\rangle \langle i~j~k~l\rangle}
\end{aligned}
\end{equation*}
where $\langle i~j~k~l\rangle$ are the minors (or Plucker coordinates) of the
matrix representing the 4-plane $Z$ and $n$ is the total number of the particles. With
the above relations we can translate any known amplitude from spinor $\lambda$ and $\lambdat$ into the new variables. 

The $\mathcal{S}$ variables can be useful in many cases. In this note we
show that using the $\mathcal{S}$ variables the recent formula
\cite{Hodges:2012ym} for tree level MHV gravity amplitudes can be made independent of reference spinors.
More precisely, MHV amplitudes in \cite{Hodges:2012ym} are computed from the determinant of a matrix $\phi_i^j$
in which the diagonal terms $\phi_i^i$ are defined as the soft factor of gravity
amplitudes. The form is simple and compact however each $\phi_i^i$ is defined with $2$ reference spinors. The formula is
independent of the choice of reference spinors only under the constraints of
momentum conservation. Using the $\mathcal{S}$ variables it is simple to remove the
dependence on reference particles.

\section{Invitation from Momentum Twistor}
\label{ch:invitation}
Before starting, let us clarify the notation.
We denote the antisymmetric contraction of $k$
elements in $\mathbb{C}^k$ by
\begin{equation*}
(\alpha_1~\alpha_2~\ldots~\alpha_k):=\epsilon_{a_1a_2\ldots a_k}
\alpha_1^{a_1}\alpha_2^{a_2}\ldots\alpha_k^{a_k}.
\end{equation*}

We start by changing the geometric structure of momentum twistors.
In momentum twistors, the intersection of two lines always gives rise to a degenerate
$2\times 2$ matrix that is the massless momentum $P_{a\dot{a}}$. This is also why the
massless property is manifest. We want to keep this idea.
We have to abandon the ``polygon'' definition $P_i=X_{i+1}-X_i$, which not only
trivializes momentum conservation but also produces the fixed ordering. 
\begin{figure}[!ht]
   \centering \includegraphics[scale=0.65]{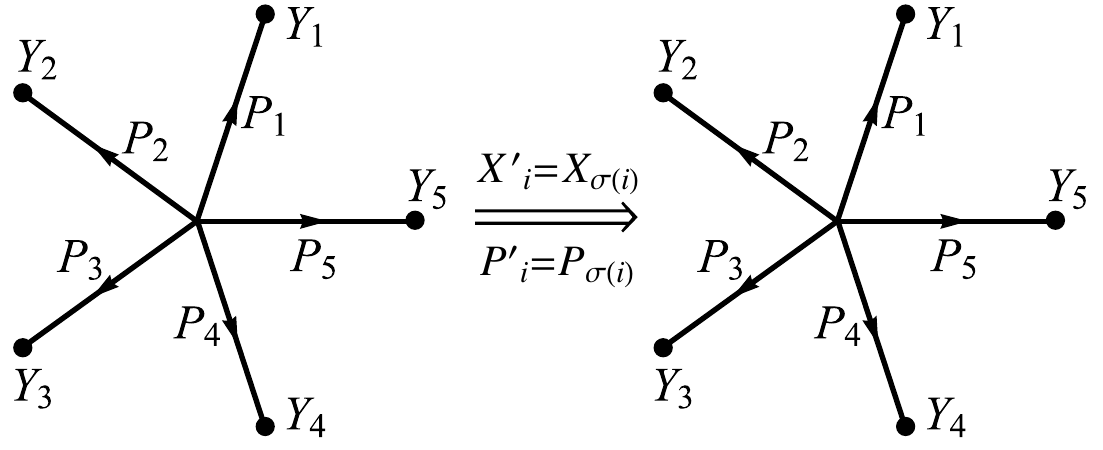}
\end{figure}
To be more
precise, we want to find another definition of $P_i$ which is also a linear
combination of lines $Y_i$ in momentum twistor space, which sum up to zero but also
stay the same while exchanging any two particle labels.
The linear combination of lines $Y_i$ satifying the requirements is the following,
\begin{equation}
  \label{eq:pmmt_star}
	P_i =Y_i-\frac{1}{n}\sum_{j=1}^n{Y_j}.
\end{equation}
Here $\frac{1}{n}\sum_{j=1}^n{Y_{ja\dot{a}}}$ is the center of mass of lines $Y_j$,
which is the same in the definition of every $P_i$.
Compared with momentum twistors, the figure above shows that performing a permutation
$\sigma$ on the points $Y_i$ and on momenta $P_i$ completely give rise to the same graph.
Therefore, we have the geometric picture in momentum twistor space.
As
indicated in \cref{eq:pmmt_star} each line $Y_i$ intersects with the center of mass
$\frac{1}{n}\sum_{j=1}^n{Y_{ja\dot{a}}}$ (the green line in the figure). 
 \begin{figure}[!ht]	
\centering \includegraphics[scale=0.7]{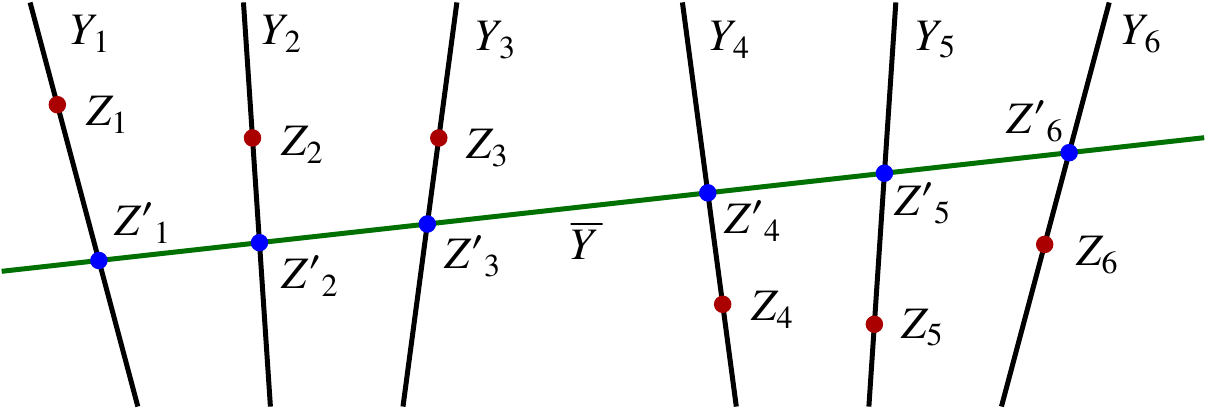}
%  \caption*{}
%  \label{fig:pmmt}
\end{figure}
\\Now we have to find a way to compute the momenta $P_i$. In
momentum twistors, $n$ points ${Z}_i$ in the twistor space uniquely give $n$ lines
$X_i$. Here we want to do something analogous. Nevertheless, we cannot simply
take the $n$ intersections points of $Y_i$ and the
center of mass line $\bar{Y}$ as the inputs. Because
here these $n$ points all lie on the same line $\bar{Y}$,
they are not independent anymore thus do not produce enough degrees of
freedom to get $Y_i$.
After a few reflections we find a proper way of definition. Each particle $i$ is
associated with one twistor ${Z}_i$ and a spinor $\lambda_i$.
\begin{equation}
  \label{eq:pmmt_input}
	{{Z}}_i =
	\left(\begin{array}{c} \zeta_i \\ \eta_i \end{array} \right),\qquad
	  {{Z}}'_i =\left(\begin{array}{c} \lambda_i \\ Y_i\lambda_i
	\end{array}\right)
\end{equation}
and the equations for $Y_i$ are
\begin{equation}
  \label{eq:pmmt_eqns}
\begin{aligned}
	 Y_{ia\dot{a}}\zeta^a_i =\eta_{i\dot{a}}, \qquad
 (Y_{ia\dot{a}}-\frac{1}{n}\sum_{j=1}^n {Y_{ja\dot{a}}})\lambda^{ia} = 0,
\end{aligned}
\end{equation}
where $\zeta_i$, $\eta_i$ and $\lambda_i$ are unconstrained inputs. The first set
of equations in \cref{eq:pmmt_eqns} means that each ${{Z}}_i$ lies on the
line $Y_i$; the second set of equation restrict each $Y_i$ intersects with the center of mass line $\bar{Y}$ at
specified $\lambda_i$. There are $4n$ equations in \cref{eq:pmmt_eqns} in total,
which are sufficient to solve the $Y_i$.
The equations might cause a little confusion. Because we know in
twistor space there is no solution for a line to intersect with more than four
arbitrary lines.
But note that in our case we do not specify the $n$ lines as inputs directly,
therefore the $n$ lines as defined are not independent and they do not apply to the
generic conclusion. In fact they are all related by the center of mass, which is again
not manually chosen but controlled by the twistors ${{Z}}_i$ and the extra spinors
$\lambda_i$.

In order to reproduce the known amplitudes by ${Z}$ and $\lambda$,
first we note that $P_{ia\dot{a}}\lambda_{ia}=0$, which means we could let
\begin{equation}
  \label{eq:pmmt_p}
\lambdat_{i\dot{a}}=\frac{P_{ia\dot{a}}}{\lambda_{ia}},
\end{equation}
where $P_{ia\dot{a}}$ comes from solving \cref{eq:pmmt_eqns} of $Y_i$. Since $(
\lambda_i~\lambda_j )$ can be obtained straightforwardly from $\lambda$. Now the goal
is to find the explicit expression of $( \lambdat_i~\lambdat_j )$ using
${Z}$ and $\lambda$.
We can write down the solutions as follows (details of the derivation can be found
in \cref{ch:solving}.),
\begin{equation}
  \label{eq:pmmt_lambdat}
(\lambdat_i~\lambdat_j) =
\frac{1}{D_n}\sum_{k,l=1}^n{(\lambda_k~\lambda_l)(
{{Z}}_i~{{Z}}_j~{{Z}}_k~{{Z}}_l)\prod^{n}_{a \neq
 i,j,k,l}(\lambda_a~\zeta_a)}
\end{equation}
in which $D_n$ is the determinant of \cref{eq:pmmt_eqns}.
We can massage \cref{eq:pmmt_lambdat} to be
\begin{align}
  \label{eq:makeup}
  (\lambdat_i~\lambdat_j)& =
\left(\frac{1}{D_n}\prod^{n}_{a=1}{(\lambda_a~\zeta_a)}\right)\sum_{k,l=1}^n{(\lambda_k~\lambda_l)
\frac{( {{Z}}_i~{{Z}}_j~{{Z}}_k~{{Z}}_l)}{(\lambda_i~\zeta_i)(\lambda_j~\zeta_j)
(\lambda_k~\zeta_k)(\lambda_l~\zeta_l)}}
\end{align}
Splitting the expression in \cref{eq:makeup}, the factor outside the summation do not
carry any indices and can be considered as an overall scaling to ${Z}$.
Therefore we define $\widehat{{Z}}$ as,
\begin{equation}
  \widehat{{Z}}_i:=\left(\frac{1}{D_n}\prod^{n}_{a=1}{(\lambda_a~\zeta_a)}\right)^\frac{1}{4}\frac{{{Z}}_i}{(\lambda_i~\zeta_i)}
\end{equation}
and \cref{eq:pmmt_lambdat} becomes
\begin{equation}
    \label{eq:hint}
(\lambdat_i~\lambdat_j) =
\sum_{k,l=1}^n{(\lambda_k~\lambda_l)(
\widehat{{Z}}_i~\widehat{{Z}}_j~\widehat{{Z}}_k~\widehat{{Z}}_l)}
\end{equation}
which is much simpler. Now we have
a permutation manifest version of momentum twistors, however the
map between $\widehat{{Z}}$ and the unconstrained ${Z}$ is a bit complicated.

\section{Definition of ${\cal S}$ Variables}
Note that \eqref{eq:hint} suggests a simpler definition. Now we directly start
from the form of \eqref{eq:hint} and define our new variables, the
$\mathcal{S}$ variables.

From \cite{ArkaniHamed:2009si}
we know that the spinor-helicity variables $\lambda$ and $\lambdat$ of
$n$ massless particles can be considered as two $2$-dimensional planes in $\mathbb{C}^n$,
\begin{equation}
  \label{eq:2plane}
  \lambda=\left(
\begin{array}{cccccc}
 \lambda ^1_1 & \lambda ^1_2 & \lambda ^1_3 & \ldots  & \lambda _{n-1}^1  & \lambda
 _n^1
 \\
 \lambda _1^2 & \lambda _2^2 & \lambda _3^2 & \ldots  & \lambda _{n-1}^2  & \lambda
 _n^2
 \\
\end{array}
\right),\qquad \lambdat=
  \left(
\begin{array}{cccccc}
 \lambdat _1^1 & \lambdat _2^1 & \lambdat _3^1 & \ldots  & \lambdat _{n-1}^1  &
 \lambdat _n^1
 \\
 \lambdat _1^2 & \lambdat _2^2 & \lambdat _3^2 & \ldots  & \lambdat _{n-1}^2  &
 \lambdat _n^2
 \\
\end{array}
\right),
\end{equation}
which are orthogonal to each other. $\lambda_i$ and $\lambdat_i$ become columns in
\cref{eq:2plane} which is the matrix representation of 2-planes.
In the $\mathcal{S}$ variables
we have the same $2$-plane $\lambda$ but instead of $\lambdat$, the 2-plane orthogonal
to the $\lambda$ plane we have an arbitrary $4$-plane $Z$,
\begin{equation}
    \label{eq:4plane}
    \lambda=\left(
\begin{array}{cccccc}
 \lambda ^1_1 & \lambda ^1_2 & \lambda ^1_3 & \ldots  & \lambda _{n-1}^1  & \lambda
 _n^1
 \\
 \lambda _1^2 & \lambda _2^2 & \lambda _3^2 & \ldots  & \lambda _{n-1}^2  & \lambda
 _n^2
 \\
\end{array}
\right),\qquad
 Z= \left(
\begin{array}{cccccc}
 Z _1^1 & Z _2^1 & Z _3^1 & \ldots  & Z _{n-1}^1  & Z
 _n^1
 \\
 Z _1^2 & Z _2^2 & Z _3^2 & \ldots  & Z _{n-1}^2  & Z
 _n^2
 \\ Z _1^3 & Z _2^3 & Z _3^3 & \ldots  & Z _{n-1}^3  & Z
 _n^3
 \\
 Z _1^4 & Z _2^4 & Z _3^4 & \ldots  & Z _{n-1}^4  & Z
 _n^4
 \\
\end{array}
\right).
\end{equation}
And similarly each particle $i$ is associated with $\lambda_i$ and $Z_i$ which
are again the columns in \cref{eq:4plane}. So the
kinematic building blocks $\langle i~j\rangle$ stay the same
\begin{equation}
\begin{aligned}
  \langle i~j\rangle &=( \lambda_i~\lambda_j ).
\end{aligned}
\end{equation}
which are the minors of $\lambda$ plane. The other kinematic
building blocks $[i~j]$ take the form of \eqref{eq:hint},
\begin{equation}
    \label{eq:main}
\begin{aligned}
[i~j]= \sum_{k,l=1}^{n}{\langle
k~l\rangle \langle i~j~k~l\rangle},
\end{aligned}
\end{equation}
in which $\langle i~j~k~l\rangle$ are the minors of the 4-plane $Z$,
\begin{equation}
\begin{aligned}
\langle i~j~k~l\rangle :=(
{Z}_i~{Z}_j~{Z}_k~{Z}_l).
\end{aligned}
\end{equation}
It can be directly proven that momentum conservation is still trivialized by
\cref{eq:main} as follows,
\begin{equation}
  \label{eq:sv_mc}
  \begin{split}
  \sum_{k=1}^{n}{ \langle  i~k  \rangle   [k~j]}  &=\sum_{k,l,m=1}^{n}{
    \langle i~k \rangle \langle l~m \rangle
  \langle k~j~l~m \rangle}\\
  &=\frac{1}{3}\sum_{k,l,m=1}^{n}{
   \left( \langle i~k \rangle \langle l~m \rangle + \langle i~l
  \rangle \langle m~k \rangle + \langle i~m \rangle \langle k~l \rangle
  \right)\langle k~j~l~m \rangle}=0,
  \end{split}
\end{equation}
by shuffling $k$, $l$ and $m$ and noticing that 
$\langle i~a \rangle \langle b~c \rangle + \langle i~b \rangle \langle c~a
  \rangle + \langle i~c \rangle \langle a~b \rangle=0$ is nothing but Schouten
  identitiy\footnote{Note that in \cref{eq:sv_mc} we have only used the antisymmetry
  but not the full Schouten identity of $\langle
  i~j~k~l\rangle$.}.

\begin{figure}[ht!]
  \centering
  \subfloat{\includegraphics[scale=0.42]{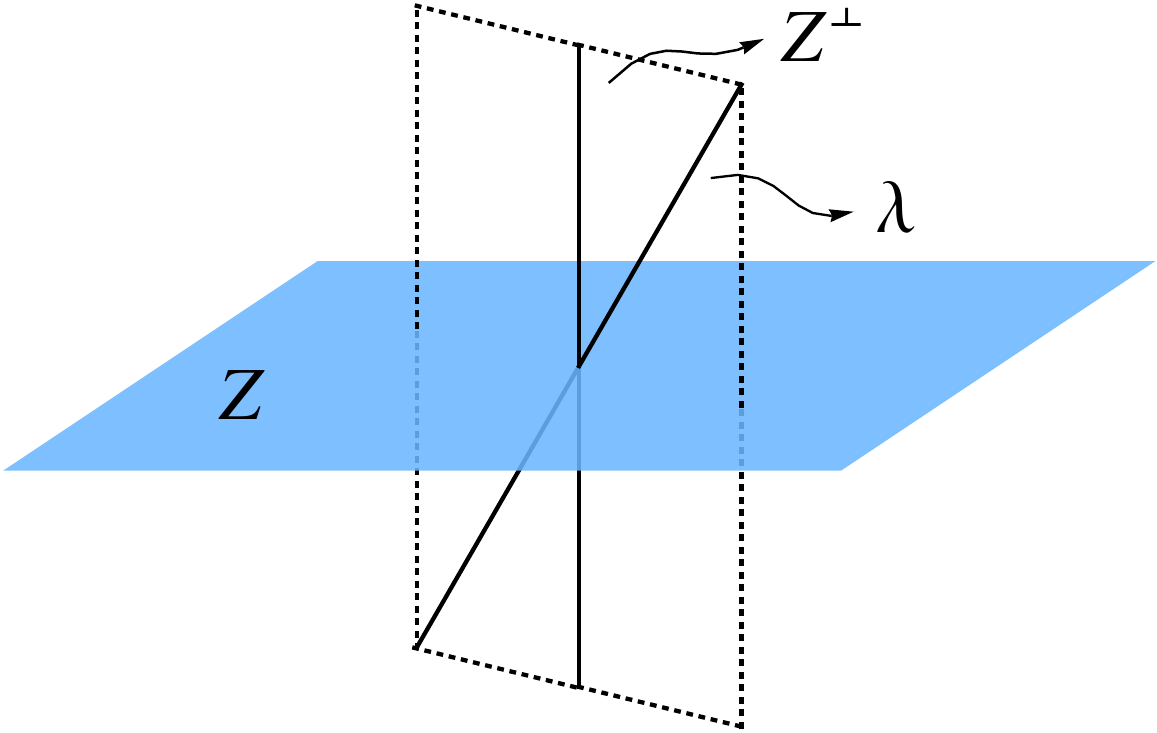}}
  \subfloat{\includegraphics[scale=0.42]{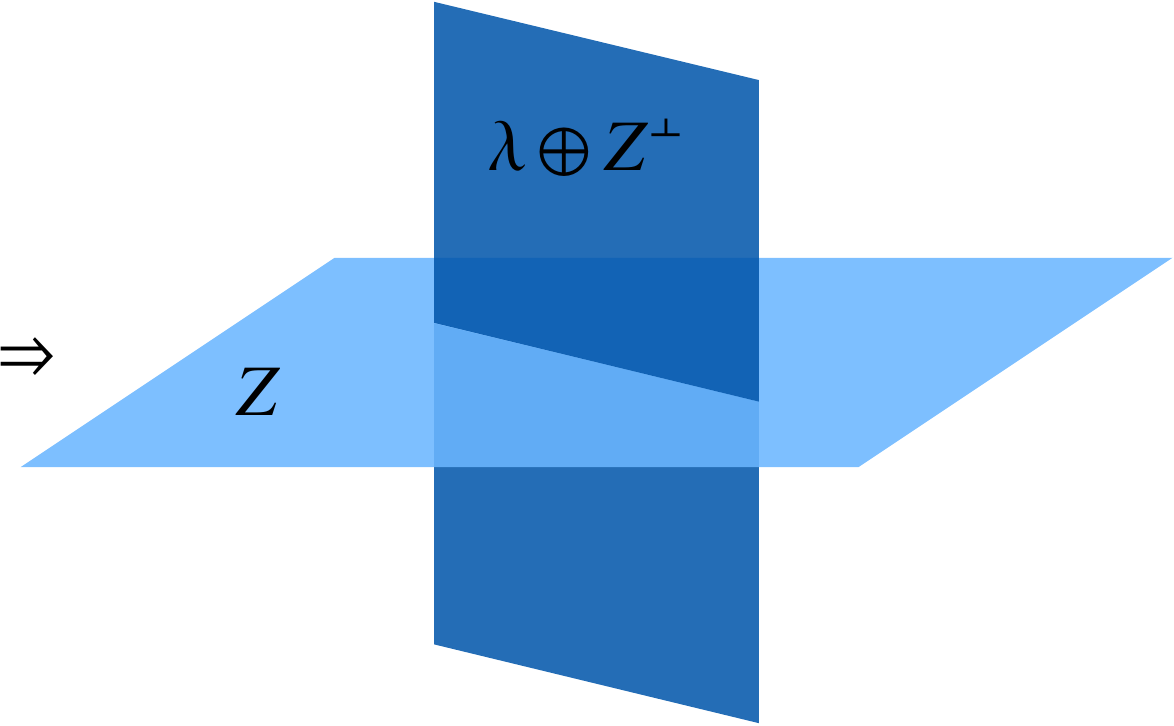}}
  \subfloat{\includegraphics[scale=0.42]{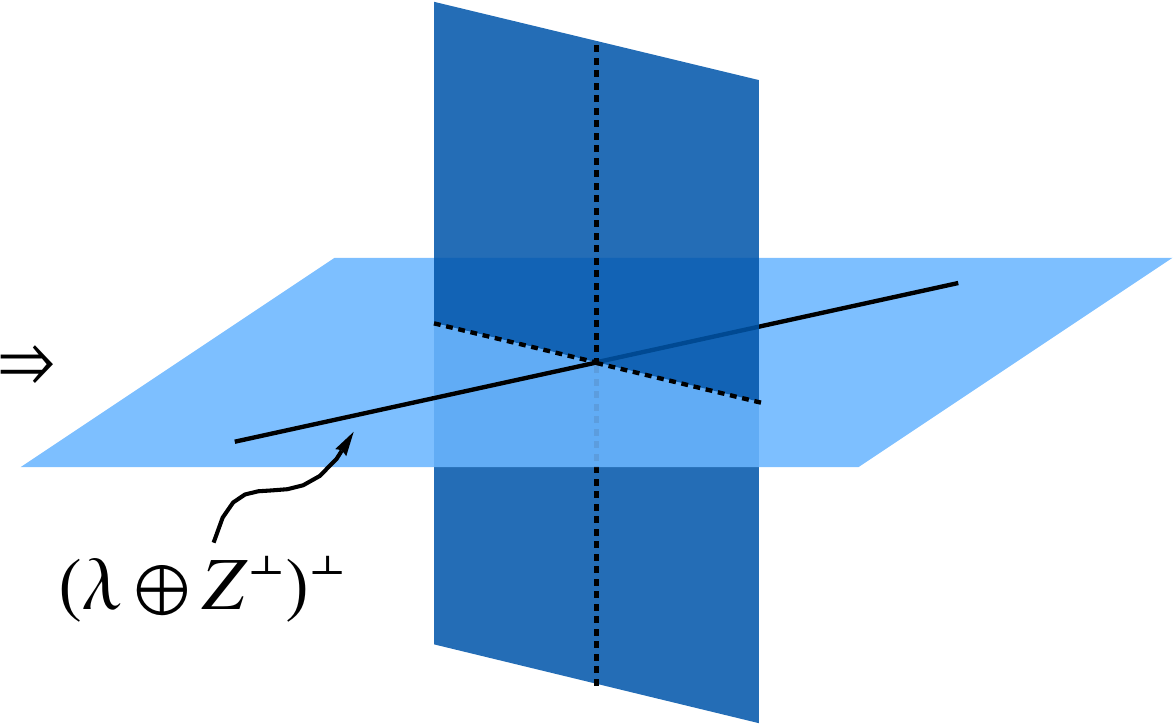}}
% \caption*{} \label{fig:f}
\end{figure}
There is a simple geometric understanding of \cref{eq:main}. Geometrically the whole
problem is to find a 2-plane $\lambdat$ to be orthogonal to the 2-plane $\lambda$ in
$\mathbb{C}^n$. One way to proceed is to start with a generic 4-plane ${Z}$ and
realize that
\begin{equation}
  \text{dim}(\lambda^\perp \cap Z)=2,
%   \qquad (\lambda^\perp \cap Z)\subset   \lambda^\perp
\end{equation}
where $\lambda^{\perp}$ is the $n-2$ dimensional plane which is
orthogonal to $\lambda$. Therefore $\lambda^\perp \cap Z$ is a solution of
$\lambdat$.
It is also easy to find that
\begin{equation}
  \lambda^\perp \cap Z=(\lambda\oplus Z^{\perp})^\perp
\end{equation}
where $Z^{\perp}$ is the $n-4$ dimensional plane orthogonal to $Z$. One can directly
prove that the \cref{eq:main} is simply the Plucker coordinate expression of
$\lambdat=(\lambda\oplus Z^{\perp})^\perp$.

In fact, $\mathcal{S}$ variables can be generalized mathematically, although it might
not be physically relevant. The general problem\footnote{Here we abuse the
notations $\lambda$, $\lambdat$ and $Z$ to denote general planes in $\mathbb{C}^n$.} is to find 
a generic $i$-dimensional plane $\lambdat$
orthogonal to a $j$-dimensional plane $\lambda$. And \cref{eq:main} can be
generalized as follows
\begin{equation}
  \label{eq:generalized}
(\lambdat_{a_1}~\lambdat_{a_2}~\ldots~\lambdat_{a_i}) \sim
\sum_{b_1,b_2,\ldots,b_j=1}^{n}{(\lambda_{b_1}~\lambda_{b_2}~\ldots~\lambda_{b_j})
({Z}_{a_1}~{Z}_{a_2}~\ldots~{Z}_{a_i}~{Z}_{b_1}~{Z}_{b_2}~\ldots~{Z}_{b_j})},
\end{equation}
which gives a solution to the problem.
Here $Z$ is a generic
$(i+j)$-dimensional plane. A brief proof of \cref{eq:generalized} is shown in
\cref{ch:proof}.

\begin{table}[ht!]
\label{tb:gaugecounting}
\centering
\begin{tabular}{|c|c|c|c|c|}
 \hline
 \text{    } & \text{\,\;Momentum\;\,} & \text{\,\;\quad Spinor \quad\;\,} &
 $\begin{array}{c} \text{Momentum}\\\text{Twistor}\end{array}$ &
 \;\;$\mathcal{S}$ Variables\;\; \\\hline
$\begin{array}{c}\text{Gauge}\\\text{Degrees of Freedom}\end{array}$ & 0 & $n$ &
$n+4$ & $3n+4$ \\ \hline $\begin{array}{c}\text{Total}\\\text{Degrees of Freedom} \end{array}$ &
$3n-4$ & $4n-4$ & $4n$ & $6n$ \\\hline $\begin{array}{c}
 \text{Number of}\\\text{Constraints}
 \end{array}$ & $n+4$ & $4$ & \cellcolor[gray]{0.9}$0$ & \cellcolor[gray]{0.9}$0$
 \\\hline
 $\begin{array}{c}
 \text{Massless}\\\text{Manifest}
 \end{array}$ & No & \cellcolor[gray]{0.9}{Yes} & \cellcolor[gray]{0.9}{Yes} &
 \cellcolor[gray]{0.9}{Yes}
 \\\hline
 $\begin{array}{c}
 \text{Momentum Conservation} \\\text{Manifest}
 \end{array}$ & No & No & \cellcolor[gray]{0.9}{Yes} &
 \cellcolor[gray]{0.9}{Yes}
 \\\hline
 $\begin{array}{c}
 \text{Permutation}\\\text{Manifest}
 \end{array}$ & \cellcolor[gray]{0.9}{Yes} & \cellcolor[gray]{0.9}{Yes} & No &
 \cellcolor[gray]{0.9}{Yes} \\\hline
\end{tabular}
\end{table}
Let us now do a simple counting of degrees of freedom to close the section. The
physical degrees of freedom of $n$ massless particles is $3n-4$. We could regard a
twistor, not in a projective way, but as a 4-component vector with one rescaling gauge
degree of freedom. In this way momentum twistors have $n+4$ gauge degrees of freedom,
in which the $4$ comes from the fact that a constant translation to each $X_i$ does not change the
kinematics. Similarly the spinors $\lambda$ and $\lambdat$ also have
rescaling gauge. For the $\mathcal{S}$ variables in
general\eqref{eq:generalized}, all $Z$ that satisfy the following form
\begin{equation}
  Z=\left(
\begin{array}{cccccc}
 1 & \cdots  & 0 & 0 & 0 & 0 \\
 \vdots  & \cdots  & \vdots  & \vdots  & \cdots  & \vdots  \\
 0 & \cdots  & 1 & 0 & 0 & 0 \\
 d^1_1 & \cdots  & d_1^i & 1 & \cdots  & 0 \\
 \vdots  & \cdots  & \vdots  & \vdots  & \cdots  & \vdots  \\
 d_j^1 & \cdots  & d_j^i & 0 & \cdots  & 1
\end{array}
\right)\left(
\begin{array}{cccccccc}
 c_1^1 & c_1^2 & \cdots  & \cdots  & \cdots  & \cdots  & c_1^{n-1} & c_1^n \\
 \vdots  & \vdots  & \cdots  & \cdots  & \cdots  & \cdots  & \vdots  & \vdots  \\
 c_i^1 & c_i^2 & \cdots  & \cdots  & \cdots  & \cdots  & c_i^{n-1} & c_i^n \\
 \lambdat _1^1 & \lambdat _1^2 & \cdots  & \cdots  & \cdots  & \cdots  & \lambdat _1^{n-1} & \lambdat _1^n \\
 \vdots  & \vdots  & \cdots  & \cdots  & \cdots  & \cdots  & \vdots  & \vdots  \\
 \lambdat _j^1 & \lambdat _j^2 & \cdots  & \cdots  & \cdots  & \cdots  & \lambdat
 _j^{n-1} & \lambdat _j^n
\end{array}
\right)
\end{equation}
give the same solution $\lambdat$, in which $c$ and $d$ are arbitrary.
Therefore the gauge degrees of freedom is rescaling plus all free
parameters $c$ and $d$, and their number equals $n+in+ij$. In particular, here when
$\lambda$ and $\lambdat$ are 2-component spinors, the number is $3n+4$.
We compare all of them
including normal 4-component momentum in the table.
So we clearly see the ``evolution'' of variables by introducing more gauge, which
makes more physical properties manifest.

\section{MHV Amplitudes of Gravity and Soft Factors}
As an application of the $\mathcal{S}$ variables, let us consider MHV amplitudes of
gravity, which is first obtained in \cite{Berends:1988zp}. Here we strip out the
momentum conserving delta function and its superpartners. 
The result for the 4 particles amplitude looks very nice in the $\mathcal{S}$
variables,
\begin{equation}
M_4= \left(\prod^4_{a<b} \langle a~b\rangle\right)^{-1}\langle1~2~3~4\rangle.
\end{equation}
The denominator is a product of all non-vanishing angular brackets. As we can see, the
result is manifestly permutation invariant.
For $n=5$, we get
\begin{equation}
  \label{eq:m5}
M_5=\left(\prod^5_{a<b} \langle a~b\rangle\right)^{-1}
\sum_{s<t,\,i<j<k}^5{\langle s~t\rangle \langle i~j\rangle \langle
j~k \rangle  \langle i~k \rangle \overline{\langle
s~i~j~k\rangle}\,\overline{\langle
t~i~j~k\rangle}},
\end{equation}
which is again manifestly permutation invariant.
The notation $\overline{\langle i~j~k~l \rangle}$ means
$i,j,k,l$ are forced to be ordered so that it has the right sign,
\begin{equation}
  \overline{\langle
i~j~k~l\rangle} =\langle a~b~c~d \rangle, \qquad a,b,c,d \in
\{i,j,k,l\},\;\;\;a<b<c<d.
\end{equation} 
For general $n$, we use the recent result\cite{Hodges:2012ym} of Hodges that
\begin{equation}
  M_n=(-1)^{n+1}\text{sgn}(\alpha\beta)c_{\alpha(1)\alpha(2)\alpha(3)}c^{\beta(1)\beta(2)\beta(3)}
  \phi_{[\alpha(4)}^{\beta(4)}\phi_{\alpha(5)}^{\beta(5)}\ldots\phi_{\alpha(n)]}^{\beta(n)}.
\end{equation} 
Here $\phi_i^j$ is an $n\times n$ matrix. Using our $\mathcal{S}$
variables, $\phi_i^j$ become the follows
\begin{equation}
    \label{eq:phi}
  \begin{aligned}
\phi_i^j&=\sum_{k,l=1}^{n}{\frac{\langle
k~l\rangle}{\langle
i~j\rangle} \langle i~j~k~l\rangle},\qquad i\neq j \\
	\phi_i^i&=\frac{1}{6}\sum_{j,k,l=1}^{n}{\frac{\langle
	j~k\rangle \langle k~l\rangle \langle l~j\rangle}
	{\langle i~j\rangle \langle i~k\rangle \langle i~l\rangle}\langle i~j~k~l\rangle}.
\end{aligned}
\end{equation}
In \cref{eq:phi} the
second formula is the soft factor for gravity amplitudes.
The permutation invariant property is
manifest for all $\phi_i^j$ and particularly the reference particles disappear in the
expression of $\phi_i^i$. In this sense the $\mathcal{S}$ variables improve
the formula for tree level MHV gravity amplitudes in \cite{Hodges:2012ym}.

\section{Discussions}
A very remarkable property of momentum twistors is that they make the dual conformal symmetry\cite{Drummond:2008vq}
manifest in planar Yang-Mills theory.
One of the motivations of this work is the hope that the $\mathcal{S}$ variables will help make properties of gravity manifest. Although there seem to be nothing like dual conformal symmetry, $\mathcal{N}=8$
supergravity is known to have an $E_{7(7)}$ symmetry. 

Another hope is that the $\mathcal{S}$ variables could be also useful understanding
the KLT\cite{Kawai:1985xq} and BCJ\cite{Bern:2008qj} relations because those
relations also depend on the constraints of momentum conservation. 

It would also be interesting to find the
$\mathcal{S}$ variables form of BCFW\cite{Britto:2005fq}. For this purpose it is
necessary to understand factorization and we give the first steps in
\cref{ch:factorization}.

\section*{Acknowledgements}
We are grateful to Freddy Cachazo for pointing out the relation of our work to
Grassmannians and many inspiring and explanatory conversations.

\appendix
% \section*{Appendix}
\section{Solving Equations}
\label{ch:solving}
We want to solve \cref{eq:pmmt_eqns}
and find a nice expression of $(\lambdat_i~\lambdat_j)$. 
Here by nice expression, we expect every
single factor could be expressed as some kinds of contractions. There are several
possible contractions allowed by the little group properties, namely,
\begin{equation}
(\lambda_i~\lambda_j),\quad ( \lambda_i~\eta_j),\quad (\lambda_i~\zeta_j),
\quad ( \zeta_i~\zeta_j),\quad ( \eta_i~\eta_j),\quad ( \zeta_i~\eta_j),
\quad ({{Z}}_i~{{Z}}_j~{{Z}}_k~{{Z}}_l ).
\end{equation}
Note $({{Z}}_i~{{Z}}_j~{{Z}}_k~{{Z}}_l )$ is not
independent but an antisymmetric sum of products of $( \zeta_i~\zeta_j)$ and $(
\eta_i~ \eta_j)$. It is unlikely that all of these contraction will appear inevitably
in the expression, so we now try to find a few clues to eliminate a few of them.

We start by writing down the equations in a matrix form. Note that $Y_{ia\dot{a}}$ can
actually be independently split into two sets: $Y_{ia1}$ and $Y_{ia2}$. Thus the $4n$
equations can be diagonalized into two sets of independent $2n$ equations with the
same array of the coefficients.
\begin{equation}
  \label{eq:pmmt_matrix}
\left(
\begin{array}{cccccc}
 \zeta _1^1 & \zeta _1^2 & 0 & 0 & \cdots  & \cdots  \\
 0 & 0 & \zeta _2^1 & \zeta _2^2 & \cdots  & \cdots  \\
 \vdots  & \vdots  & \vdots  & \vdots  & \cdots  & \cdots  \\
 \frac{n-1}{n}\lambda _1^1 & \frac{n-1}{n}\lambda _1^2 & -\frac{1}{n}\lambda _1^1 &
 -\frac{1}{n}\lambda _1^2 & \cdots  & \cdots  \\
 -\frac{1}{n}\lambda _2^1 & -\frac{1}{n}\lambda _2^2 & \frac{n-1}{n}\lambda _2^1 &
 \frac{n-1}{n}\lambda _2^2 & \cdots  & \cdots  \\
 \vdots  & \vdots  & \vdots  & \vdots  & \cdots  & \cdots
\end{array}
\right)\left(
\begin{array}{c}
 Y_{111} \\
 Y_{121} \\
 Y_{211} \\
 Y_{221} \\
 \vdots  \\
 \vdots
\end{array}
\right)=\left(
\begin{array}{c}
 \eta_{11}  \\
 \eta_{21}  \\
 \vdots  \\
 0 \\
 0 \\
 \vdots
\end{array}
\right)
\end{equation}
\begin{equation}
\left(
\begin{array}{cccccc}
 \zeta _1^1 & \zeta _1^2 & 0 & 0 & \cdots  & \cdots  \\
 0 & 0 & \zeta _2^1 & \zeta _2^2 & \cdots  & \cdots  \\
 \vdots  & \vdots  & \vdots  & \vdots  & \cdots  & \cdots  \\
 \frac{n-1}{n}\lambda _1^1 & \frac{n-1}{n}\lambda _1^2 & -\frac{1}{n}\lambda _1^1 &
 -\frac{1}{n}\lambda _1^2 & \cdots  & \cdots  \\
 -\frac{1}{n}\lambda _2^1 & -\frac{1}{n}\lambda _2^2 & \frac{n-1}{n}\lambda _2^1 &
 \frac{n-1}{n}\lambda _2^2 & \cdots  & \cdots  \\
 \vdots  & \vdots  & \vdots  & \vdots  & \cdots  & \cdots
\end{array}
\right)\left(
\begin{array}{c}
 Y_{112} \\
 Y_{122} \\
 Y_{212} \\
 Y_{222} \\
 \vdots  \\
 \vdots
\end{array}
\right)=\left(
\begin{array}{c}
 \eta_{12}  \\
 \eta_{22}  \\
 \vdots  \\
 0 \\
 0 \\
 \vdots
\end{array}
\right)
\end{equation}
Following Cramer's formulas, the solutions of $Y_{ia\dot{a}}$ must share the
determinant of the coefficient array of the linear equations as a common denominator
$D_n$. We get
\begin{equation}
  \label{eq:pmmt_deno}
D_n:=-\frac{1}{n^n}\sum_{\sigma\in
S_n}^{}{(1-n)^{a(\sigma)}\text{sgn}(\sigma)\prod_{i=1}^{n}{(\lambda_i~\zeta_{\sigma(i)})}}.
\end{equation}
The numerator is the difficult part. We find it by induction. For the 4
particles case,
\begin{equation}
  \label{eq:ket1234}
(\lambdat_1~\lambdat_2) =
\frac{1}{D_4}(\lambda_3~\lambda_4)({{Z}}_1~{{Z}}_2~{{Z}}_3~
{{Z}}_4).
\end{equation}
Eq.~\eqref{eq:ket1234} gives us some hint of the structure of the numerator. First,
from $({{Z}}_1~{{Z}}_2~{{Z}}_3~{{Z}}_4)$, we can rule out the standalone
contraction $( \eta_i~\eta_j)$, because we know from \cref{eq:pmmt_matrix} that
$(\lambdat_1~\lambdat_2)$ must only be degree 2 in $\eta$, which is already in
$({{Z}}_1~{{Z}}_2~{{Z}}_3~{{Z}}_4)$. We also know
from \cref{eq:pmmt_deno} that the denominator has $\lambda$ of degree $n$ and $\zeta$
of degree $n$. Together with \cref{eq:pmmt_p}, we conclude that the numerator must
have $\lambda$ of degree $n-2$, $\zeta$ of degree $n-2$ and $\eta$ of degree 2.

Now we make a guess based on the above analysis of degrees and \cref{eq:ket1234}.
We assume that, for generic $n$, the numerator of $(\lambdat_i~\lambdat_j)$ has the
contraction $({{Z}}_a~{{Z}}_b~{{Z}}_c~{{Z}}_d)$ and
$(\lambda_e~\lambda_f)$, which left with $\lambda$ of degree $n-4$, $\zeta$ of
degree $n-4$. It is reasonable to guess that
they form a degree $n-4$ polynomial in $(\lambda_g~\zeta_h)$. The $n-4$ also coincides
with the fact that we do not see this monomial in $n=4$ case.
Indeed, for $n=5$, we get
\begin{multline}
(\lambdat_1~\lambdat_2) =
\frac{1}{D_5}[(\lambda_5~\zeta_5)(\lambda_3~\lambda_4)(
{{Z}}_1~{{Z}}_2~{{Z}}_3~
{{Z}}_4) \\ +(\lambda_4~\zeta_4)(\lambda_3~\lambda_5)(
{{Z}}_1~ {{Z}}_2~ {{Z}}_3~
{{Z}}_3)+(\lambda_3~\zeta_3)(\lambda_4~\lambda_5)(
{{Z}}_1~{{Z}}_2~{{Z}}_4~
{{Z}}_5)]
\end{multline}
Now it is easy to conjecture the formula to be,
\begin{equation}
  \label{eq:raw}
(\lambdat_i~\lambdat_j) =
\frac{1}{D_n}\sum_{k,l=1}^n{(\lambda_k~\lambda_l)(
{{Z}}_i~{{Z}}_j~{{Z}}_k~{{Z}}_l)\prod^{n}_{a \neq
 i,j,k,l}(\lambda_a~\zeta_a)}
\end{equation}
and we have checked numerically that it is correct.

We end the section by discussion of direct gauge fixing for \cref{eq:makeup}. Recall
our degree counting in \cref{tb:gaugecounting}, we have the rescaling gauge for each
${Z}_i$ and also for $\lambda_i$ so we can choose a gauge for $\zeta_i$ and
$\lambda_i$ as follows,
 \begin{equation}
   \label{eq:nicegauge}
   \zeta_i=\left(\begin{array}{c}1\\\alpha_i\end{array}\right),\qquad
   \lambda_i=\left(\begin{array}{c}\frac{\beta_i}{\alpha_i}\\\beta_i+1\end{array}\right)
\end{equation}
It is clear both $\zeta_i$ and $\lambda_i$ can run over the projective $\mathbb{CP}^1$
independently except several singularities. Under this gauge, each
\begin{equation}
  (\lambda_i~\zeta_i)=1, \qquad i=1,2,\ldots,n
\end{equation}
And \cref{eq:raw} literally becomes
\begin{equation}
(\lambdat_i,\lambdat_j) =
\frac{1}{D_n}\sum_{k,l=1}^n{(\lambda_k~\lambda_l)(
{{Z}}_i~{{Z}}_j~{{Z}}_k~{{Z}}_l)}.
\end{equation}

\section{Proof of the Generalized Formula}
\label{ch:proof}
We split the proof into two steps. First we list a proposition. And we
use the proposition to prove the corollary which is \cref{eq:generalized}.

\noindent{\bf Proposition. }
% {\em
$V_i$ are $k$-component vectors and
$W_i$ are $(n-k)$-component vectors, $i=1,2,\ldots,n$.
$V$ is a $k$-plane and $W$ is a $(n-k)$-plane in $\mathbb{C}^n$ that
\begin{equation}
  V=(V_{1},V_{2},\ldots,V_{n}), \quad W=(W_{1},W_{2},\ldots,W_{n}),\quad V\perp W.
\end{equation}
Then $\exists$
constant $\alpha\neq 0$,
\begin{equation}
  \label{eq:proposition}
  (V_{a_1}~V_{a_2}~\cdots~V_{a_k})=\frac{\text{sgn}(\sigma)}{\alpha}
  (W_{\hat{a}_1}~W_{\hat{a}_2}~\cdots~W_{\hat{a}_{n-k}})
\end{equation}
for any $a_1$,  $a_2$, \ldots, $a_k$ and $\hat{a}_1$, $\hat{a}_2$,
\ldots, $\hat{a}_{n-k}$. Here $\sigma$ is a permutation that
\begin{equation}
  \sigma=\left(\begin{array}{cccccccc}
    1 & 2 & \ldots & k & k+1 & k+2 & \ldots & n \\
  a_1 & a_2 & \ldots  & a_k & \hat{a}_1 & \hat{a}_2 & \ldots & \hat{a}_{n-k}
  \end{array}
  \right).
\end{equation}

\noindent{\bf Corollary.}
$\lambdat_l$ are $i$-component vectors,
$\lambda_l$ are $j$-component vectors and
$Z_l$ are $(i+j)$-component vectors, $l=1,2,\ldots,n$.
$\lambdat$ is a $i$-plane, $\lambda$ is a $j$-plane and $Z$ is a $(i+j)$-plane in
$\mathbb{C}^n$ that
\begin{equation}
  \lambdat=(\lambdat_{1},\lambdat_{2},\ldots,\lambdat_{n}),\quad
  \lambda=(\lambda_{1},\lambda_{2},\ldots,\lambda_{n}), \quad
  Z=(Z_{1},Z_{2},\ldots,Z_{n}).
\end{equation}
If $\lambdat=(\lambda\oplus Z^{\perp})^\perp$ then $\exists$
constant $\alpha\neq 0$,
\begin{equation}
(\lambdat_{a_1}~\lambdat_{a_2}~\ldots~\lambdat_{a_i})
=
\frac{1}{\alpha}\sum_{b_1,b_2,\ldots,b_j=1}^{n}
{(\lambda_{b_1}~\lambda_{b_2}~\ldots~\lambda_{b_j})
({Z}_{a_1}~{Z}_{a_2}~\ldots~{Z}_{a_i}~{Z}_{b_1}~{Z}_{b_2}~\ldots~{Z}_{b_j})}
\end{equation}
for any $a_1$,  $a_2$, \ldots, $a_i$ and $b_1$, $b_2$,
\ldots, $b_j$. 

\noindent{\bfseries\em Proof.}
Let us first define a $k$-plane $X$ and a $(n-i)$-plane $W$ to be
\begin{equation}
    X=Z^{\perp},\qquad W=\lambda\oplus Z^{\perp}
\end{equation}
where $k=n-i-j$.
We have the matrix representation of $W$,
\begin{equation}
  W=\left(
\begin{array}{cccccccc}
 \lambda_1^1 & \lambda_1^2 & \cdots  & \cdots  & \cdots  & \cdots  & \lambda_1^{n-1} &
 \lambda_1^n \\
 \vdots  & \vdots  & \cdots  & \cdots  & \cdots  & \cdots  & \vdots  & \vdots  \\
 \lambda_j^1 & \lambda_j^2 & \cdots  & \cdots  & \cdots  & \cdots  & \lambda_j^{n-1} &
 \lambda_j^n \\
 X_1^1 & X_1^2 & \cdots  & \cdots  & \cdots  & \cdots  & X_1^{n-1} & X_1^n \\
 \vdots  & \vdots  & \cdots  & \cdots  & \cdots  & \cdots  & \vdots  & \vdots  \\
 X_{k}^1 & X_{k}^2 & \cdots  & \cdots  & \cdots  & \cdots  & X
 _{k}^{n-1} & X_{k}^n
\end{array}
\right).
\end{equation}
Using the \cref{eq:proposition}, we have
\begin{equation}
    \label{eq:step1}
  (\lambdat_{a_1}~\lambdat_{a_2}~\ldots~\lambdat_{a_i})=\frac{1}{\alpha_1}\text{sgn}(\sigma_1)
  (W_{\hat{a}_1}~W_{\hat{a}_2}~\cdots~W_{\hat{a}_{n-i}}),
\end{equation}
where $\sigma_1$ is a permutation that
\begin{equation}
  \sigma_1=\left(\begin{array}{cccccccc}
    1 & 2 & \ldots & i & i+1 & i+2 & \ldots & n \\
  a_1 & a_2 & \ldots  & a_i & \hat{a}_1 & \hat{a}_2 & \ldots & \hat{a}_{n-i}
  \end{array}
  \right).
\end{equation}
We can see
\begin{equation}
    \label{eq:step2}
  (W_{\hat{a}_1}~W_{\hat{a}_2}~\cdots~W_{\hat{a}_{n-i}})=\sum_{\sigma\in
  S_{n-i}} {
  \text{sgn}(\sigma)(\lambda_{{b_1}}~\lambda_{{b_2}}~\ldots~\lambda_{{b_j}})
  ({X}_{c_1}~{X}_{c_2}~\ldots~{X}_{c_{k}})},
\end{equation}
where 
\begin{equation}
  \begin{aligned}
  b_1=\sigma(\hat{a}_1),\quad
  b_2=\sigma(\hat{a}_2),\quad\ldots, \quad b_j=\sigma(\hat{a}_j),\\
  c_1=\sigma(\hat{a}_{j+1}),\quad
  c_2=\sigma(\hat{a}_{j+2}),\quad \ldots, \quad c_k=\sigma(\hat{a}_{n-i}),
\end{aligned}
\end{equation}
and $S_{n-i}$ is the set of all permutations
of $\{\hat{a}_1,\hat{a}_2,\ldots,\hat{a}_{n-i}\}$. 
Using the \cref{eq:proposition} again, we have
\begin{equation}
  \label{eq:step3}
({X}_{c_1}~{X}_{c_2}~\ldots~{X}_{c_{k}})=
 \frac{1}{\alpha_2} \text{sgn}(\sigma_2)
({Z}_{a_1}~{Z}_{a_2}~\ldots~{Z}_{a_i}~{Z}_{b_1}~{Z}_{b_2}~\ldots~{Z}_{b_j}),
\end{equation}
$\sigma_2$ is a permutation that
\begin{equation}
  \sigma_2=\left(\begin{array}{cccccccccccc}
    1 & 2 & \ldots & k & k+1 & k+2 & \ldots & n-j & n-j+1 & n-j+2 & \ldots
    & n
    \\
  c_1 & c_2 & \ldots  & c_{k} & a_1 & a_2 & \ldots & a_i & b_1 & b_2 & \ldots  &
  b_j
  \end{array}
  \right).
\end{equation}
With \cref{eq:step1,eq:step2,eq:step3}, we have
\begin{equation}
(\lambdat_{a_1}~\lambdat_{a_2}~\ldots~\lambdat_{a_i})
=
\frac{1}{\alpha}\sum_{\sigma
\in S_{n-i}}{(\lambda_{b_1}~\lambda_{b_2}~\ldots~\lambda_{b_j})
({Z}_{a_1}~{Z}_{a_2}~\ldots~{Z}_{a_i}~{Z}_{b_1}~{Z}_{b_2}~\ldots~{Z}_{b_j})}.
\end{equation}
Here we
have
\begin{equation}
  \frac{1}{\alpha}=(-1)^{(i+j)(n-i-j)}\text{sgn}(\sigma_1)\frac{1}{\alpha_1}\frac{1}{\alpha_2},
\end{equation}
% $\frac{1}{\alpha}=(-1)^{(i+j)(k)}\text{sgn}(\sigma_1)\frac{1}{\alpha_1}\frac{1}{\alpha_2}$,
Because determinants are antisymmetric, we can let
\begin{equation}
  \sum_{\sigma\in S_{n-i}}\rightarrow \sum_{b_1,b_2,\ldots,b_j=1}^{n},
\end{equation}
which finishes the proof.

\section{Geometry of Factorization}
\label{ch:factorization}
Factorization arises when the sum of momenta of a subset of particles become massless.
We can put an on-shell propagator between the two subsets of particles.
Kinematics of the two subsets becomes independent after cutting the propagator.
We define the division of the two subsets to be $L$ and $R$ with
\begin{equation}
  L\cup R=\{1,2,\ldots,n-1,n\} \nonumber
\end{equation}
and two lines $Y_L$ and $Y_R$ related to the propagator,
\begin{equation*}
\begin{aligned}
Y_L-\bar{Y}&=\sum_{i\in{L}}{(Y_i-\bar{Y})} \\
Y_R-\bar{Y}&=\sum_{i\in{R}}{(Y_i-\bar{Y})}.
\end{aligned}
\end{equation*}
Take the 6 particles case as an example. Say $L=\{1,2,3\}$ and $R=\{4,5,6\}$, and we
have
\begin{align}
Y_L&=Y_1+Y_2+Y_3-2 \bar{Y} \nonumber\\
Y_R&=Y_4+Y_5+Y_6-2 \bar{Y} \nonumber
\end{align}
The lines $Y_L$ and $Y_R$ only intersect with $\bar{Y}$ when
factorization arises, as shown in the
figure.
\begin{figure}[!ht]
		\centering
	\includegraphics[scale=0.7]{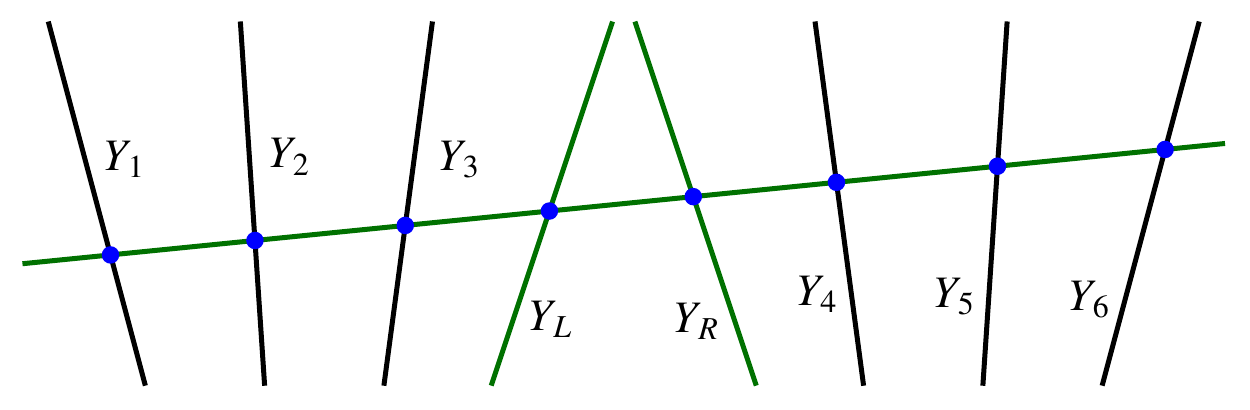}
%     \caption{}\label{fig:fac_pmmt_2}
\end{figure}
And $\{Y_1,Y_2,Y_3,Y_R\}$ and $\{Y_4,Y_5,Y_6,Y_L\}$ both become momentum
conserved by themselves. The center of mass line of both subsets is still $\bar{Y}$.
\begin{equation*}
  \frac{1}{4}(Y_R+Y_1+Y_2+Y_3)=\frac{1}{4}(Y_L+Y_4+Y_5+Y_6)=\bar{Y}
\end{equation*}


\begin{thebibliography}{9}

%\cite{Xu:1986xb}
\bibitem{Xu:1986xb}
  Z.~Xu, D.~-H.~Zhang and L.~Chang,
  ``Helicity Amplitudes for Multiple Bremsstrahlung in Massless Nonabelian Gauge Theories,''
  Nucl.\ Phys.\ B {\bf 291}, 392 (1987).
  %%CITATION = NUPHA,B291,392;%%

%\cite{Hodges:2009hk}
\bibitem{Hodges:2009hk}
  A.~Hodges,
``Eliminating spurious poles from gauge-theoretic amplitudes,''
  arXiv:0905.1473 [hep-th].
  %%CITATION = ARXIV:0905.1473;%%
  
%\cite{Alday:2007hr}
\bibitem{Alday:2007hr} 
  L.~F.~Alday and J.~M.~Maldacena,
  ``Gluon scattering amplitudes at strong coupling,''
  JHEP {\bf 0706}, 064 (2007)
  [arXiv:0705.0303 [hep-th]].
  %%CITATION = ARXIV:0705.0303;%%

%\cite{ArkaniHamed:2009si}
\bibitem{ArkaniHamed:2009si}
  N.~Arkani-Hamed, F.~Cachazo, C.~Cheung and J.~Kaplan,
  ``The S-Matrix in Twistor Space,''
  JHEP {\bf 1003}, 110 (2010)
  [arXiv:0903.2110 [hep-th]].
  %%CITATION = ARXIV:0903.2110;%%

%\cite{Hodges:2012ym}
\bibitem{Hodges:2012ym}
  A.~Hodges,
  ``A simple formula for gravitational MHV amplitudes,''
  arXiv:1204.1930 [hep-th].
  %%CITATION = ARXIV:1204.1930;%%

%\cite{Berends:1988zp}
\bibitem{Berends:1988zp}
  F.~A.~Berends, W.~T.~Giele and H.~Kuijf,
  ``On relations between multi - gluon and multigraviton scattering,''
  Phys.\ Lett.\ B {\bf 211}, 91 (1988).
  %%CITATION = PHLTA,B211,91;%%

%\cite{Drummond:2008vq}
\bibitem{Drummond:2008vq}
  J.~M.~Drummond, J.~Henn, G.~P.~Korchemsky and E.~Sokatchev,
  ``Dual superconformal symmetry of scattering amplitudes in N=4 super-Yang-Mills theory,''
  Nucl.\ Phys.\ B {\bf 828}, 317 (2010)
  [arXiv:0807.1095 [hep-th]].
  %%CITATION = ARXIV:0807.1095;%%
  
%\cite{Kawai:1985xq}
\bibitem{Kawai:1985xq} 
  H.~Kawai, D.~C.~Lewellen and S.~H.~H.~Tye,
  ``A Relation Between Tree Amplitudes of Closed and Open Strings,''
  Nucl.\ Phys.\ B {\bf 269}, 1 (1986).
  %%CITATION = NUPHA,B269,1;%%

%\cite{Bern:2008qj}
\bibitem{Bern:2008qj}
  Z.~Bern, J.~J.~M.~Carrasco and H.~Johansson,
  ``New Relations for Gauge-Theory Amplitudes,''
  Phys.\ Rev.\ D {\bf 78}, 085011 (2008)
  [arXiv:0805.3993 [hep-ph]].
  %%CITATION = ARXIV:0805.3993;%%
  
%\cite{Britto:2005fq}
\bibitem{Britto:2005fq} 
  R.~Britto, F.~Cachazo, B.~Feng and E.~Witten,
  ``Direct proof of tree-level recursion relation in Yang-Mills theory,''
  Phys.\ Rev.\ Lett.\  {\bf 94}, 181602 (2005)
  [hep-th/0501052].
  %%CITATION = HEP-TH/0501052;%%

\end{thebibliography}
\end{document}